# Energy dissipation mechanism revealed by spatially resolved Raman thermometry of graphene/hexagonal boron nitride heterostructure devices

Daehee Kim[1,2†], Hanul Kim[3†], Wan Soo Yun[2], Kenji Watanabe[4], Takashi Taniguchi[4], Heesuk Rho[3*], Myung-Ho Bae[1,5*]

[1]Korea Research Institute of Standards and Science, Daejeon 34113, Republic of Korea

[2]Department of Chemistry, Sungkyunkwan University, Suwon 16419, Republic of Korea

[3]Department of Physics, Research Institute of Physics and Chemistry, Chonbuk National University, Jeonju 54896, Republic of Korea

[4]National Institute for Materials Science, Tsukuba, Ibaraki, 305-0044, Japan

[5]Department of Nano Science, University of Science and Technology, Daejeon 34113, Republic of Korea

[†]These authors contributed equally to this work.

[*]Contacts: rho@chonbuk.ac.kr, mhbae@kriss.re.kr

## Abstract

Understanding the energy transport by charge carriers and phonons in two-dimensional (2D) van der Waals heterostructures is essential for the development of future energy-efficient 2D nano-electronics. Here, we performed in situ spatially resolved Raman thermometry on an electrically biased graphene channel and its hBN substrate to study the energy dissipation mechanism in graphene/hBN heterostructures. By comparing the temperature profile along the biased graphene channel with that along the hBN substrate, we found that the thermal boundary resistance between the graphene and hBN was in the range of $(1 - 2) \times 10^{-7}$ $m^2 KW^{-1}$ from ~100 °C to the onset of graphene break-down at ~600 °C in air. Consideration of an electro-thermal transport model together with the Raman thermometry conducted in air showed that local doping occurred under a strong electric field played a crucial role in the energy dissipation of the graphene/hBN device up to $T \sim 600$ °C.

**Introduction**

The development of two-dimensional (2D) heterostructures composed of graphene, hexagonal boron nitride (hBN) and transition metal dichalcogenides (TMDCs) has opened a new avenue beyond the current Si-based electronics. The presence of van der Waals (vdW) interactions between layers of 2D atomic crystals can be used in designing novel electronic, optoelectronic and photonic devices by enabling the manufacturer to control the thickness of the device on the atomic scale and to stack different 2D crystals in diverse ways.[1,2,3,4,5] To fully understand the operating mechanism of devices made of vdW heterostructures, it is crucial to study how heat is generated and dissipated in these devices, and how such generation and dissipation are affected by the transport of hot charge carriers and phonons through the different 2D layers. For example, the energy-relaxation process of photo-excited hot carriers has been studied in various vdW heterostructures such as $MoS_2/WS_2$, graphene/hBN/graphene and graphene/semiconducting TMDC/graphene.[6,7,8] Furthermore, the phonon transport through the vdW heterostructures has been studied in terms of the thermal boundary resistance (TBR) between dissimilar 2D materials.[9,10]

To develop energy-efficient 2D electronics, both electric transport and energy transport should be considered simultaneously in order to understand more comprehensively the role of the energy dissipation in the vdW heterostructures. While heat management has been extensively studied in conventional Si devices, III-V semiconductor-based heterostructures[11] and graphene (or MoS$_2$)/SiO$_2$ devices,[12,13] the study of the mechanism of energy dissipation in 2D vdW heterostructures is still in its infancy. For a graphene/hBN device, for example, since the power density per unit length along the graphene channel under a high bias is not uniform, one needs to measure temperature ($T$) profiles of both graphene and hBN together with carrying out proper numerical electro-thermal transport calculations in order to fully investigate the genuine role of energy dissipation in the vdW-heterostructure devices.

In this study, we performed spatially resolved Raman thermometry on biased graphene/hBN heterostructure devices in air to investigate the energy dissipation in the vdW heterostructure (see Figure 1a). The measured temperature of the hBN ($T_{BN}$) underneath an operating graphene field-effect transistor (FET) with a channel length ($L$) of 26 μm agreed well with the $T_{BN}$ estimated by an analytical calculation for $T_{BN} \leq 100$ °C. Furthermore, to explore power dissipation in an extreme case, we carried out studies in which we approached the break-down $T$ of graphene (~600 °C) in air with a relatively short, $L = 5$ μm, graphene device. By comparing the temperature profile along a biased graphene channel ($T_{GR}$) with that along the hBN substrate, we found $R_{GB}$ to be in the range (1 - 2) × 10$^{-7}$ m$^2$KW$^{-1}$ for $T = 100$ - 600 °C. Also, by performing Raman thermometry together with numerical calculations using an electro-thermal transport model, we showed that the local doping under a high-field transport limit significantly affected the energy dissipation in the graphene channel on hBN for the entire $T$ range examined.

**Results and Discussions**

To realize a graphene device on an hBN layer, we placed mechanically exfoliated hBN flakes on top of a 90-nm-thick SiO$_2$/Si substrate. Then, onto a selected 80-nm-thick hBN flake (Supporting Information, Figure S1), we placed a single layer of graphene using a transfer method.[14] We then used electron-beam lithography to deposit metal electrodes (Cr (5 nm)/Au (95 nm)) for the source and drain. For the electrical measurements, the highly $p$-doped Si substrate was used as an electrode to apply a back-gate voltage ($V_{bg}$). Figure 1b shows an optical microscope image of the completed device with a channel length ($L$) of 26 μm and width ($W$) of 6 μm, where 'S' and 'D' mark the locations of the source and drain, respectively. The mobility of charge carriers in the graphene was measured to be ~7000 cm$^2$V$^{-1}$s$^{-1}$ based on the fit of the resistance *vs*. $V_{bg}$ curve (Supporting Information, section 1). For the Raman thermometry, we performed the $T$ calibration process using a laser beam with an excitation wavelength of 633 nm.[9,15] For this device, the intensities of the G and 2D peaks were not sufficiently strong to define the energy values through the graphene channel (Supporting Information, Figure S3). Thus, we obtained the experimental $T_{BN}$ map from the hBN substrate based on the $E_{2g}$ phonon energy of hBN beneath the graphene channel for a case of this device. The dependence of the $E_{2g}$ phonon energy of hBN on $T$ was determined by heating the graphene/hBN FET device on a hot plate, which gave a slope, $S_{BN}$, of $-0.0218$ cm$^{-1}$/°C (Supporting Information, Figure S4a). We also confirmed that the 633-nm- wavelength laser did not result in any significant photo-induced doping of the graphene channel on hBN[16] (Supporting Information, section 2). In all of hBN thermometry experiments, we did not apply a finite $V_{bg}$.

Figure 1c shows the representative Raman spectra of the hBN underneath the graphene FET at $V_{sd}$ = 0, 10, $-10$, 15 and $-15$ V, which were recorded at the middle of the channel. The dashed vertical line indicates the location of the $E_{2g}$ phonon energy at $V_{sd}$ = 0 V. For finite $V_{sd}$, the $E_{2g}$ phonon energy showed a red-shift behavior. For example, the $E_{2g}$ phonon energy shifted downward from 1366.22 to 1364.55 cm$^{-1}$ when $V_{sd}$ was changed from zero to $-15$ V, where the peak energies were obtained from the Lorentzian fit as shown by the red curves on the Raman spectra. From the value of $S_{BN}$ indicated above and the 22.4 °C value of the ambient $T$, $T_0$, we estimated the $T_{BN}$ at the middle of the channel to be ~100 °C at $V_{sd} = -15$ V based on a relation of $T = T_0 + (\omega_{BN} - \omega_{BN}^0)/S_{BN}$, where $\omega_{BN}^0$ and $\omega_{BN}$ are the $E_{2g}$ energies of the hBN at zero and finite $V_{sd}$, respectively. Based on this process, we produced maps, with pixel dimensions of $1 \times 1$ μm$^2$, of the $T_{BN}$ underneath the graphene channel for various $V_{sd}$ values (Figure 1d) by scanning the laser beam over the entire graphene channel. Figure S5 in Supporting Information shows $E_{2g}$-energy maps of the hBN underneath the graphene channel at corresponding $V_{sd}$ values including the zero bias condition. The average $T$ of the entire hBN area underneath the graphene channel was found to increase with increasing $V_{sd}$, i.e., to 43.2 °C (47.7 °C) and 76.4 °C (92.3 °C) at $V_{sd} = 10$ (−10) and 15 (−15) V, respectively. Interestingly, the left side was hotter than the right side with a difference of ~30 °C at $V_{sd} = 15$ V, in contrast to a nearly homogeneous $T$ distribution through the entire hBN channel at $V_{sd} = -15$ V, and this issue is discussed below.

For a detailed analysis, we plotted, as shown in Figure 2a-d, profiles of $T_{BN}$ (triangles or solid circles) along the hBN beneath the graphene channel at $V_{sd} = 10$, −10, 15 and −15 V, respectively, where the $T_{BN}$ shown at each indicated $x$ position was calculated as the average of the $T$ values over the $y$ axis in Figure 1d. The average $T_{BN}$ showed relatively similar values of ~45 °C and ~50 °C for $V_{sd} = 10$ V and −10 V (see Figure 2a,b), respectively, but rather dissimilar values of ~81 °C and ~96 °C for $V_{sd} = 15$ and −15 V (see Figure 2c,d), respectively. To gain insight into the measured $T$ values for graphene and hBN, we analytically calculated $T$ values based on the schematic shown in Figure 2e. In the steady state, the average $T$ of the graphene ($T_{AG}$) for a given electrical power ($P$) was calculated from the total thermal resistance ($R_{tot}$) from the graphene to the heat sink of the Si substrate according to the equation

$$T_{AG} = R_{tot}P + T_0 \tag{1}$$

, where $R_{\text{tot}} = \frac{R_{\text{GB}}}{LW} + R_{BN} + \frac{R_{\text{BO}}}{LW} + R_{\text{ox}} + \frac{R_{\text{OS}}}{LW} + R_{\text{Si}}$. $R_{\text{GB}}$ (~$1.35 \times 10^{-7}$ m$^2$KW$^{-1}$) [9], $R_{\text{BO}}$ (~$2.2 \times 10^{-8}$ m$^2$KW$^{-1}$),[17] and $R_{\text{OS}}$ (~$10^{-8}$ m$^2$KW$^{-1}$) [18] are the TBRs between graphene and hBN, hBN and SiO$_2$, and SiO$_2$ and Si, respectively. $R_{\text{BN}}$ ($= t_{BN}/(k_{BN}^{\perp}LW)$), $R_{\text{ox}}$ ($= t_{\text{ox}}/(k_{\text{ox}}LW)$), and $R_{\text{Si}}$ ($= 1/(2k_{\text{Si}}\sqrt{LW})$) are the thermal resistances of the hBN, SiO$_2$ and Si substrate, respectively.[12] Here, $k_{\text{BN}}^{\perp}$ (~3 Wm$^{-1}$K$^{-1}$), $k_{\text{ox}}$ (~1.3 Wm$^{-1}$K$^{-1}$) and $k_{\text{Si}}$ (~50 Wm$^{-1}$K$^{-1}$) are the thermal conductivities of hBN (out-of-plane),[19] SiO$_2$ and the highly doped Si, respectively.[12] We obtained $R_{\text{tot}}$ ~ 2480 K/W, resulting in a $T_{AG}$ of ~ 63.8 °C with a $P$ ($= IV_{\text{sd}} - I^2R_{\text{s}}$) of ~ 16.7 mW at $V_{\text{sd}} = \pm 10$ V, and with $R_{\text{s}}$ (= 950 Ω) being the series resistance. At $V_{\text{sd}} = \pm 15$ V with a $P$ of ~ 38 mW, we obtained a $T_{AG}$ of ~ 116 °C. Then, we estimated the $T$ drop through the interface between graphene and hBN, using the relationship $\Delta T_{\text{GB}} = PR_{\text{GB}}/(LW)$, to be ~14.5 and ~32.8 °C, resulting in ~ 49 and ~83 °C values for $T_{BN}$, at $V_{\text{sd}} = \pm 10$ V and ±15 V, respectively. The analytically calculated $T_{BN}$ values for $V_{sd} = 10, -10$ and 15 V, as shown by the dashed horizontal black lines in Figure 2a-c, respectively, were found to be consistent with the corresponding measured $T_{\text{BN}}$ values (solid triangles or solid circles) as shown by the dashed horizontal black lines in these figures.

The averaged measured $T$ values of hBN at $V_{\text{sd}} = -15$ V, however, as shown by the solid circles in Figure 2d were found to be as much as ~13 °C greater than the analytically estimated values (dashed horizontal line). We attributed this difference to a local doping under the condition of high-field transport with a relatively high $V_{\text{sd}}$ value. For graphene on a SiO$_2$ substrate, a locally high electric field has been shown to induce trapped charges in SiO$_2$ near the interface between graphene and SiO$_2$, and to hence result in the local doping[20]. The local doping was itself be manifested by the appearance of another new local resistance maximum aside from the original local resistance maximum observed at the charge neutral gate voltage ($V_{\text{g0}}$) in the $R$-$V_{\text{bg}}$ curves. In the case of graphene on hBN, there was only one local resistance maximum at $V_{\text{g0}}$ ~ 13 V in the beginning of the thermometry experiment as shown in Figure 3a. After applying a $V_{sd}$ of 10 V for the Raman thermometry, however, an additional local resistance maximum at $V_{bg} \sim 0$ V appeared (see a blue curve in Figure 3b). Then, it disappeared after applying a $V_{sd}$ of $-10$ V and was again observed after applying a $V_{sd}$ of 15 V (see blue curves in Figure 3c,d). Thus, we concluded that there was a local doping caused by the locally high field in the graphene on the hBN layer, as has previously been observed for graphene on $SiO_2$ layer.

The $-10$ μm $< x < 3$ μm region of the channel was measured to be hotter than the other regions when the $V_{sd}$ was 15 V, as shown in Figure 1d. The greater $T$ of this region reflects the electric field of the corresponding graphene region being stronger than that of the other regions[12], which may have led to a high-field-induced doping effect in this left (or drain) side of the graphene on hBN. Furthermore, the Joule-heat-induced high $T$ likely resulted in local annealing, which could have led to the additional local doping in the graphene. Such local doping leads to different Fermi energy levels along the graphene channel, as shown in the upper panel of Figure 2f with $V_{sd} = 0$ V, where the dashed line indicates the Fermi energy level. The portion of channel near the drain has less hole doping than that near the source, according to the observation of a newly appeared local resistance maximum in the $R$-$V_{bg}$ curve located near $V_{bg} \sim 0$ V together with the maintenance of the original local resistance maximum at $V_{bg} \sim 13$ V (see Figure 3d). With the non-uniform carrier density profile along the graphene channel as an initial condition, applying a $V_{sd}$ of $-15$ V led to a change in the Fermi level along the graphene channel, as shown in the lower panel in Figure 2f. Interestingly, in that case, the carrier density resided near the charge-neutral condition along the entire graphene channel, in contrast to the case for which a $V_{sd}$ of 15 V was applied (see the lower panel of Figure 2f). This resulted in a relatively uniform and high resistance through the entire channel, and thus a higher average experimentally determined $T$ (solid circles in Figure 2d) than expected from the calculations (dashed horizontal line in Figure 2d).

The analytical approach based on eq 1 assumed that the graphene channel was exposed to a uniform electric field or homogeneous power density. But, the electric field would be expected to vary along the channel under a high-$V_{sd}$ regime due to the carrier density significantly varying along the channel (see the lower panel of Figure 2f). Thus, to reveal the genuine $T$ profile of the graphene channel and the local field strength causing local doping at graphene on hBN, we performed a numerical simulation based on the electro-thermal transport model. We numerically calculated the $T$ profile along the graphene and hBN by considering the local electric field $F_x$ as given by the equation $F_x = -dV_x/dx$, where $V_x$ is the local voltage in the graphene channel with $x = -L/2 \sim L/2$ [12]. The local $T$ ($T_x$) along a channel of interest with a local power density per unit length $p'_x$, calculated using the equation $p'_x = IF_x$, and a current $I$ is given by the steady-state heat equation

$$Ak\frac{d^2T_x}{dx^2} + p'_x - \frac{1}{LR^{\perp}_{\mathrm{th}}}(T_x - T_0) = 0 \qquad (2)$$

, where $A$ is the cross-sectional area of the heat channel, $k$ is the in-plane thermal conductivity of the heat channel and $R^{\perp}_{\mathrm{th}}$ is the thermal resistance from the material of interest to the heat sink. In the case of graphene, the $F_x$ was calculated by the drift velocity and mobility with local carrier density based on $R$-$V_{\mathrm{bg}}$ curves (with the calculation details provided elsewhere[12]). Figures 3d and 3e show $R$-$V_{bg}$ curves (blue) after thermometry performed at $V_{sd} = 15$ V and $-15$ V, respectively. For the numerical calculations of $T$ profiles, we used red curves of the fits to get mobility and $V_{g0}$ in corresponding data. We also obtained the voltage ($V_x$) drop along the graphene by integrating $F_x$ along the channel length. In this way, we calculated $I$-$V_{sd}$ curves (solid curves), which are shown along with the experimental data (scattered points) in Figure S6 in Supporting Information; the $I$-$V_{sd}$ curves were obtained while the bias voltage reached to $V_{sd} = \pm 10$ V and $\pm 15$ V.

Figure 2a-d show the $T$ profiles of graphene (dashed red curves) and hBN (solid red curves) obtained by carrying out simulations at $V_{sd} = 10$, $-10$, $15$, and $-15$ V, respectively. In eq 2, we considered $R_{th}^{\perp} = R_{tot}$ with $k = k_G$ (= 700 $Wm^{-1}K^{-1}$) to calculate the $T$ profile in the graphene channel on the hBN, where $k_G$ is the thermal conductivity of graphene[21] on hBN. Here, we used an $R_{GB}$ value of $1.35 \times 10^{-7}$ $m^2KW^{-1}$ for the calculations based on a recent report,[9] although the value could be lowered by improving the quality of the graphene/hBN interface.[10] The result showed a local $T$ maximum (or hot spot) in the graphene channel for all conditions, where the location of the hot spot depended on the $V_{bg}$ and $V_{sd}$ conditions. For example, a hot spot was located at the drain when a $V_{sd}$ of 15 V was used (see dashed curve in Figure 2c). The presence of a hot spot here was attributed to the local minimum in the hole carrier density at the drain in a hole-doped region (see the lower panel of Figure 2f), resulting in a local maximum drift velocity ($v_{dx}$) with a continuity equation for a current density of $j = n_x e v_{dx}$, where $n_x$ and $e$ are a local total carrier density and elementary change, respectively. This also led to a local maximum $F_x$ at the drain with the relationship $v_{dx} \sim F_x \mu$, accompanied by a local maximum power density and hot spot in that region.

On the other hand, the experimentally obtained $T$ profiles along the hBN (triangles or solid circles) in Figure 2a-d did not show such sharp hot spots as the T profiles calculated for graphene channel. This difference was related to the ratio between the heat dissipation rates through lateral and substrate directions. The length scale of the lateral heat dissipation is the lateral healing length, $L_H = L\sqrt{R_{th}^{\perp}/R^{\parallel}}$, where $R^{\parallel} \left[ = \left( k^{\parallel}A/L \right)^{-1} \right]$ is the in-plane thermal resistance of the material of interest[12]. For the hBN with in-plane thermal conductivity[22] $k_{BN}^{\parallel}$ of 400 $Wm^{-1}K^{-1}$ and $t_{BN}$ of 80 nm, we obtained an $L_{H,BN}$ (healing length of hBN layer) value of ~2.4 μm. Here, a total thermal resistance from hBN to a Si heat sink is given by $R_{th,BN}^{\perp} = R_{tot} - \frac{R_{GB}}{LW}$. In the case of the graphene, on the other hand, an $L_{H,G}$ value of ~ 0.3 μm was obtained for $k_G = 600$-$900$ Wm$^{-1}$K$^{-1}$ with a total thermal resistance from graphene to a heat sink of $R_{th,G}^{\perp} = R_{tot}$, resulting in relatively sharp hot spots as shown by the dashed lines in Figure 2a-d. For the numerical calculation of the $T$ profile along the hBN, we used $A = W \times t_{BN}$ as the cross-sectional area for the lateral heat flow. The heat flux density $p'_x$ used for hBN was the same as that calculated from the graphene channel. The calculated $T$ profiles along the hBN channel denoted by the solid red curves in Figure 2a-c showed good agreement with the experimental data for $V_{sd} = 10, -10$ and $15$ V, respectively.

At $V_{sd} = -15$ V, the experimentally obtained $T$ profile of hBN deviated significantly from the calculation (solid red curve in Figure 2d) at the drain region but not at the source region. In the electro-thermal calculation, we assumed a uniform doping with a single local resistance maximum at $V_{g0} = 12$ V (see the red curve in Figure 3e). However, the experimentally obtained $R$-$V_{bg}$ curve (blue curve in Figure 3e) showed an additional local resistance maximum (vertical arrow) at $V_{bg} \sim 3$ V, which indicated a local doping phenomenon. This local doping apparently resulted in another hot spot near the drain region, and hence led to $T$ rising in the drain as much as at the source in the hBN, as shown in Figure 2d (solid circles). This result is consistent with the analytic analysis in the previous section.

To determine the field-strength range causing the local doping, we plotted the electric-field strength ($F_x$) along the graphene channel when using a $V_{sd} = 15$ V (solid curve), as shown in Figure 3f. The region indicated to be locally doped, i.e., corresponding to the relatively high-$T$ region in the panel for $V_{sd} = 15$ V as shown in Figure 1d, is indicated by the double-headed arrow in Figure 3f; the magnitude of $F_x$ throughout this region was found to be greater than 0.4 V/μm. On the other hand, we also observed two local resistance maxima in the $R$-$V_{bg}$ curves after applying a $V_{sd}$ of 10 V, as shown in Figure 3b (blue curve). At $V_{sd}$ = 10 V, the magnitude of the calculated electric field strength (dashed curve in Figure 3f) in the drain part was determined to be about ~0.35 V/μm. These results indicated an electric field strength of ~0.35 V/μm to be sufficient to produce local doping in graphene on hBN. In the case of graphene on SiO$_2$, the local doping due to the electric field occured with higher fields of $F$ > 2 V/μm [20] than that of graphene on hBN. While further study is needed to understand the difference between the activation electric field needed to induce a local doping in graphene on hBN and that in graphene on SiO$_2$, a similar situation was also observed in photo-induced doping experiments, i.e., the charge-trap energy of ~2.6 eV was obtained for SiO$_2$,[23] larger than that for hBN (~2.3 eV).[16]

To explore the heat dissipation at an extreme case near the graphene break-down $T$ of ~ 600 °C in air,[24] we fabricated another graphene device on an 8-nm-thick hBN flake/290-nm-thick SiO$_2$/Si substrate and having $L$ = 5 μm and $W$ = 2 μm (top panel of Figure 4a), i.e., with an area an order of magnitude smaller than that of the device described in previous sections. For this smaller-area device, the Raman 2D peak of graphene was clearly observed through the entire channel (see Figure 4b and Supporting Information, Figure S7); thus, we successfully obtained the $T$ maps of graphene and hBN simultaneously. The solid black circles in Figure 4b show $E_{2g}$ and 2D modes of hBN and graphene, respectively, which were taken without bias voltage from the location of the channel indicated by the laser beam in the top panel of Figure 4a. When a bias voltage ($V_{sd}$ = 8.5 V) was applied, the $E_{2g}$ and 2D modes were shifted to lower energies (see solid red circles in Figure 4b). By following a $T$-calibration process, plots of $T$ versus the $E_{2g}$ and 2D mode energies gave slopes $S_{BN}$ = −0.0341 cm$^{-1}$/°C [25] and $S_{GR}$ = −0.0428 cm$^{-1}$/°C, respectively (Supporting Infor-mation, section 3). Based on these $S_{BN}$ and $S_{GR}$ values and data in Figure 4b (solid curves: Lorentzian fit results), we estimated the temperatures of graphene and hBN at the position indicated by the laser spot to be, respectively, ~470 °C and ~180 °C when applying a $V_{sd}$ of 8.5 V at $T_0$ = 24 °C. Based on this process, we mapped $T_{GR}$ and $T_{BN}$ at $V_{sd}$ = 8.5 V, by scanning the laser beam over the entire device channel (see middle (graphene) and bottom (hBN) panels in Figure 4a). While the hBN showed gradual changes in $T$ along the channel was changed, there was a rather marked change in $T$ at the middle of the graphene channel. For further analysis, we plotted $T$ profiles for the graphene (squares) and hBN (circles) channels in Figure 4c, where the error bars reflect the variation in $T$ along the channel width at a given $x$ location. The steep change in $T$ of graphene occurred between $x = -1$ μm and $x = 0$ μm, as indicated by grey squares; this marked change in $T$ was attributed to a local doping effect as discussed below.

In Figure 4d, the resistance changed in some regions during the course of the Raman scanning for hBN and graphene, and these change were attributed to changes in $V_{g0}$ during the scanning process (Supporting Information, Figure S8). Note that the resistance did not change with considerable time periods with two resistance values of ~6.3 kΩ and ~7.5 kΩ, respectively, marked (i) and (ii). These resistance levels are denoted by red and green points in the figure, respectively. The resistance changed with time between these regions (denoted by the grey points). The red, grey and green squares in Figure 4c for the graphene were obtained in the same colored-graphene time zones in Figure 4d, respectively. The region scanned during the time period in which the resistance increased with time (grey points after ~0.5 hour in the graphene scanning region in Figure 4d) coincided with a region showing a steep change in $T$ with change in position (grey squares between $x$ ~ -1 μm and $x$ ~ 0.5 μm for the graphene channel in Figure 4c). An $I$-$V_{sd}$ curve (scattered green dots in Figure 4e) was obtained as soon as finishing the Raman scanning. It indicated a resistance of ~7.5 k$\Omega$ at $V_{sd}$ = 8.5 V, consistent with region (ii). The solid green curve at 0 $\mu$m < $x$ < 2.5 $\mu$m for the $T_{GR}$ profile in Figure 4c is a calculation result at $V_{sd}$ = 8.5 V with the solid green $I$-$V_{sd}$ curve in Figure 4e. The calculation results are consistent with experimental data. Here, the charge-neutral voltage $V_{g0}$ was located at 5.6 V for region (ii) (see dashed green curve in Supporting Information, Figure S8b).

On the other hand, when the scanning was started from the drain region, the resistance of the graphene was included in region (i) as shown in Figure 4d. This condition gave the $T_{GR}$ profile for −2.5 $\mu$m < $x$ < −1 $\mu$m (red squares) in Figure 4c. To calculate the $T_{GR}$ profile for region (i), we first plotted its calculated $I$-$V_{sd}$ curve (as shown by a solid red curve in Figure 4e), from which we derived an $R$ of ~6.3 k$\Omega$ at $V_{sd}$ = 8.5 V. Here, $V_{g0}$ was determined to be 10.6 V for the calculation, in contrast to the 5.6 V value determined for the case of region (ii) (see a dashed red curve in Supporting Information, Figure S8b). This change in $V_{g0}$ perhaps resulted from the high-field-induced doping effect, since the local field strength was already greater than 1 V/$\mu$m through the entire channel. The calculation taking into account the doping effect yielded results found to be in good agreement with the experimental $T_{GR}$ profile for −2.5 $\mu$m < $x$ < −1 $\mu$m (solid red curve in Figure 4c). From region (i) to region (ii), i.e., from $x$ = −1 to 0.5 $\mu$m, $T_{GR}$ was observed to decrease from ~450 °C to ~350 °C (Figure 4c), due to a doping effect. This result indicated the energy dissipation in the graphene channel to be strongly affected by a doping effect near the graphene break-down $T$. We also calculated the $T_{BN}$ profile (circles: experiment; red and green curves on hBN data: calculations in Figure 4c), by considering the change in resistance between regions (i) and (ii) in Figure 4d.

Note that $R$ = 5.5 k$\Omega$ at the initial state in Figure 4d quickly changed to the region (i) as soon as the Raman scanning was started. The experimentally determined $I$-$V_{sd}$ curve (orange dots) was obtained for bias conditions up to $V_{sd} = 8.5$ V before the scanning. We calculated an $I$-$V_{sd}$ curve (a dashed orange curve in Figure 4e) and $T_{GR}$ profile (a dashed orange curve in Figure 4c) at this condition, and the $T_{GR}$ nearly reached ~600 °C. Fortunately, the resistance quickly increased and reached the region (i) before the graphene channel disintegrated. Changing the bias polarity from 8.5 V to −8.5 V left the sample resistance at ~5.5 kΩ (Supporting Information, Figure S9). At this condition, however, a doping effect did not occur and the graphene channel finally disintegrated in 50 sec. The AFM image of Figure 4a was obtained after the break-down event with $V_{sd} = -8.5$ V, and showed a graphene-broken region near the source. During the calculation, TBR values between graphene and hBN in the range of (1 - 2) ×$10^{-7}$ m$^2$KW$^{-1}$ were obtained for $T = 100$ - 600 °C, consistent with values previously obtained at a $T$ of ~ 100 °C [9]. We also plotted the $T_{GR}$ calculated with $R_{GB} = 2 \times 10^{-8}$ m$^2$KW$^{-1}$ (dashed grey curve in Figure 4c) with the condition of region (ii); this $R_{GB}$ value is the lowest $R_{GB}$ value measured by experiment so far to the best of our knowledge[10]. We found that a reduction of the TBR by one order of magnitude resulted in a reduction of the local Joule heating effect by a factor of 1.3 at the same power.

**Conclusions**

We performed in situ Raman thermometry of the graphene and hBN in graphene/hBN 2D heterostructure devices, and these results together with consideration of an electro-thermal transport model indicated that a strong electric field induced local doping and hence significantly affected how energy was dissipated in these devices. The thermal boundary resistance between graphene and hBN was determined to be in the range (1 - 2)×$10^{-7}$ m$^2$KW$^{-1}$ from 100 °C up to the graphene break-down $T$ of 600 °C in air. Although the estimated thermal boundary resistance was found to be an order of magnitude higher than that recently measured,[10] using a high-quality 2D material/hBN interface is nevertheless essential for realizing efficient heat management in future heterostructure 2D electronics, as shown by our electro-thermal transport model. We expect our Raman thermometry of hBN to be easily extended to various 2D vdW heterostructure devices based on TMDC materials and black phosphorous, which we expect to provide useful information for the design of energy-efficient 2D heterostructure devices.

## Methods

The in situ Raman measurements were taken using a backscattering geometry. Incident light from an He-Ne laser with a wavelength of 632.8 nm was focused on the graphene/hBN FET channel surface through an optical microscope objective lens ($50\times$ / 0.55NA). Scattered light was dispersed through a spectrometer equipped with a 1200-groove/mm grating and detected using a thermoelectrically cooled charge-coupled device detector. For the spatially resolved Raman measurements, the sample was placed on a computer-controlled piezoelectric stage that was moved over the entire transport channel in either 0.5 or 1 μm intervals. The data for each map were acquired over the course of about two hours. For the $T$-dependent Raman measurements, the sample was placed on a heating plate.

## Acknowledgments

We thank Dr. Bum-Kyu Kim for constructing the Raman scanning stage. This work was supported by the Korea Research Institute of Standards and Science under the auspices of the project 'Convergent Science and Technology for Measurements at the Nanoscale' (16011060), the Basic Science Research Program through the National Research Foundation of Korea (NRF) (Grant Nos. 2012-0009565, 2014R1A1A2057173, 2015R1A2A1A10056103,

SRC2016R1A5A1008184, and 2016R1D1A1B03935270) funded by the Ministry of Education, and the Korea Basic Science Institute under the R&D program (Project No. E37800) supervised by the Ministry of Science, ICT and Future Planning. This work has been also partly supported by the Korea-Hungary joint laboratory program for Nanosciences through the National Research Council of Science and Technology.

**Figures and captions**

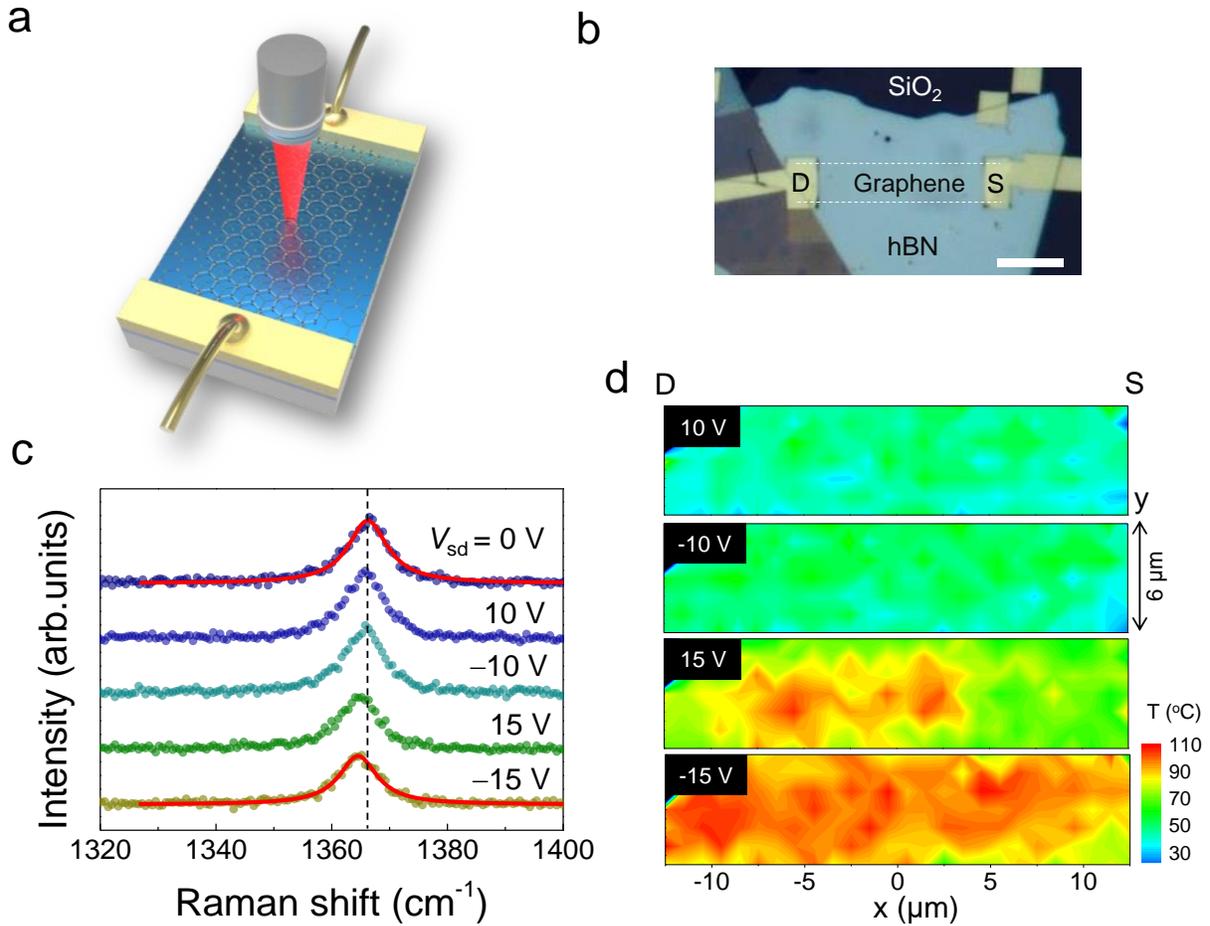

**Figure 1.** (a) Schematic of the measurement configuration in a laser scanning mode. (b) Optical microscope image of a graphene field-effect transistor ($L = 26$ µm and $W = 6$ µm) on an hBN/SiO$_2$/Si substrate. The dashed lines indicate the boundary of the graphene channel. The scale bar is 10 µm. (c) Raman spectra with the hBN $E_{2g}$ peaks at $V_{sd} = 0$, 10, -10, 15 and −15 V, which were taken at the middle region of the channel. Red curves are the Lorentzian fits. The vertical dashed line indicates the location of the peak at zero bias. (d) $T_{BN}$ maps underneath the graphene channel for various $V_{sd}$ values. The left and right ends are the drain (D) and source (S), respectively.

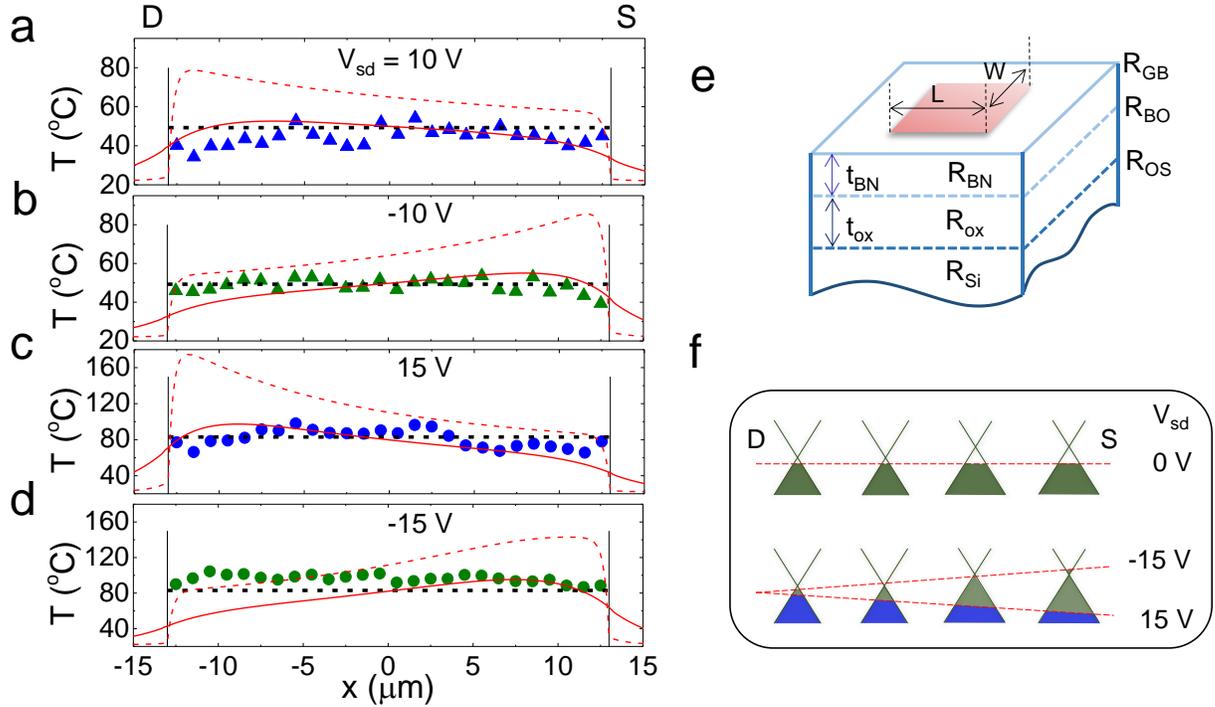

**Figure 2.** (a)-(d) $T_{BN}$ profiles (triangles or solid circles) along the hBN, where $T$ at an $x$-position was averaged through the channel width in Figure 1d. The two vertical lines indicate the boundaries of the graphene along the length of the graphene channel. Horizontal dashed lines are analytically calculated $T_{BN}$ based on the simple model shown in panel (e). Dashed and solid curves indicate the temperatures of, respectively, graphene and hBN numerically calculated based on our electro-thermal transport model. (e) Schematic of the thermal model used to analytically calculate the temperatures of graphene and hBN with a given power and thermal environment information. (f) Schematic of the energy bands (solid lines) and Fermi energy levels (dashed lines) along the graphene channel. The upper panel shows a non-uniform carrier doping along the graphene channel due to the local doping without biasing, after applying a $V_{sd}$ of 15 V. The lower panel shows the shift of the Fermi level along the graphene channel for $V_{sd} = \pm 15$ V.

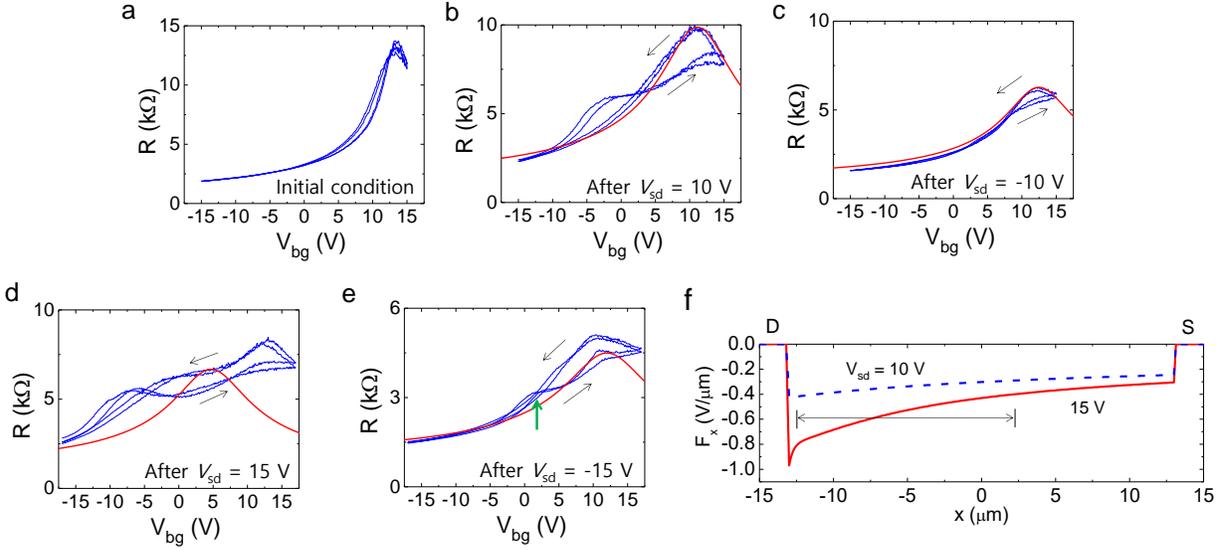

**Figure 3.** (a)-(e) $R$-$V_{bg}$ curves (blue curves) obtained following the experimental procedure after biasing various values of $V_{sd}$, where arrows indicate the sweep direction for the double sweeps. Red curves in (b)-(e) are the results of fitting to get mobility. These curves were used to calculate the $T$ profiles of the corresponding conditions shown in Figure 2a-d. To simplify the calculation of the $R$-$V_{bg}$ curve for a $V_{sd}$ of 15 V (d), we considered the fit result (red curve), which showed a charge-neutral voltage between the two observed local resistance maxima with a similar slope to them. The green arrow in (e) shows an additional local maximum resistance due to a local doping, which was already formed in the previous scanning process with $V_{sd}$ = 15 V. (f) Calculated electric-field profiles of the graphene channel for $V_{sd}$ = 10 V (dashed curve) and 15 V (solid curve), where the region spanned by the double-headed arrow is that for which local doping was expected under a $V_{sd}$ of 15 V.

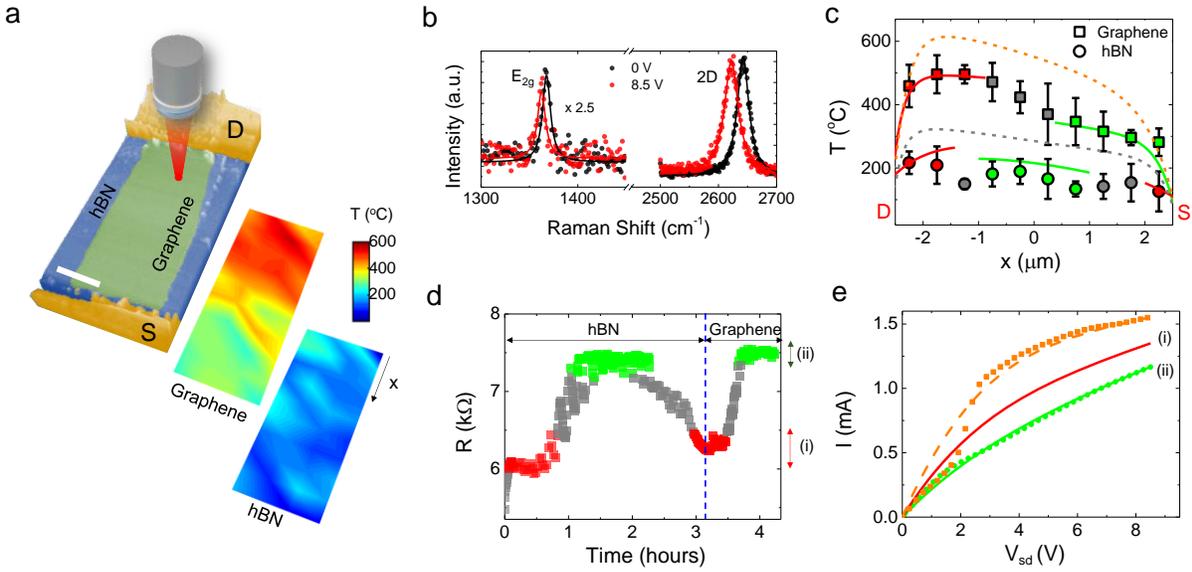

**Figure 4.** (a) Top: AFM image of the device and a schematic of the laser beam. Scale bar: 1 µm. Middle and bottom: *T* maps of graphene and hBN channels at $V_{sd} = 8.5$ V. The step size for the scanning was 0.5 µm. (b) Raman spectra taken from the region of the device indicated by the laser in (a) and showing the hBN-$E_{2g}$ and graphene-2D peaks when using a $V_{sd}$ of 0 (solid black circles) and 8.5 V (solid red circles). Black and red curves are the Lorentzian fits for $V_{sd} = 0$ and 8.5 V, respectively. (c) Squares and circles: *T* profiles of graphene and hBN channels at $V_{sd} = 8.5$ V, respectively, which were obtained from the corresponding *T* maps in (a). Here, the red, green and grey shapes correspond to temperatures obtained from regions (i) and (ii), and the region between them in (d), respectively. The solid red and green curves are results of calculations with the conditions of regions (i) and (ii) , respectively. The dashed orange curve is the result of the calculation with an initial condition of $R = 5.5$ kΩ at $V_{sd} = 8.5$ V together with the conditions of region (i). The dashed grey curve was obtained with $R_{GB} = 2 \times 10^{-8}$ m²KW⁻¹ together with the conditions of region (ii). (d) Resistance as a function of Raman thermometry scanning time values with a $V_{sd}$ of 8.5 V. (e) Scattered orange points: *I*-$V_{sd}$ curve for $V_{sd}$ values up to 8.5 V applied for the thermometry. Scattered green points: *I*-$V_{sd}$ curve just after the thermometry, and corresponding to region (ii) in (d). Red curve: *I*-$V_{sd}$ curve satisfying the condition $R = 6.3$ kΩ at $V_{sd} = 8.5$ V and corresponding to region (i)

in (d). Solid green and dashed orange curves: calculated fits to the corresponding experimentally derived

$I$-$V_{sd}$ curves.

*Supporting Information*

# Energy dissipation mechanism revealed by spatially resolved Raman thermometry of graphene/hexagonal boron nitride heterostructure devices

Daehee Kim, Hanul Kim, Wan Soo Yun, Kenji Watanabe, Takashi Taniguchi, Heesuk Rho and Myung-Ho Bae

## Section 1. Characterization of a graphene device on hBN with $L = 26$ μm

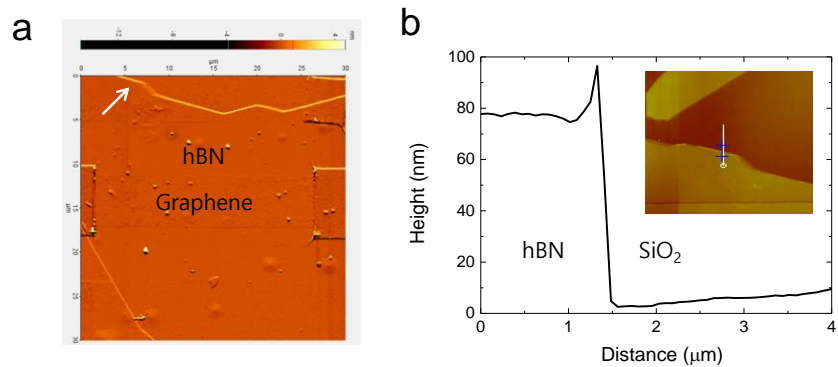

**Figure S1.** (a) AFM amplitude image of the graphene on hBN. (b) The height profile along the white line of the inset, which corresponds to a height topography of the region indicated by an arrow in (a). The thickness of the hBN is ~80 nm.

Figure S2a shows a slightly hysteretic resistance as a function of $V_{bg}$ ($R$-$V_{bg}$ curve, scattered green dots) in an ambient condition with a charge neutral gate-voltage, $V_0 \sim 13$ V. The dashed curve is a calculation result with a relation of $R = R_s + \rho_g L/W$. Here, $R_s$ (= 0.95 kΩ) is the serial resistance including two contact resistance, $\rho_g$ the sheet resistivity of the graphene, and $L$ (= 26 μm) and $W$ (= 6 μm) the length and width of the graphene device, respectively. The $\rho_g$ is given by $[q\mu(n + p)]^{-1}$, where $q$ is the elementary charge, $\mu$ is the low-field mobility and $n$ ($p$) is the electron (hole) density. The charge density is determined by the gate-induced and thermally excited carriers including the puddle density[1]. We obtained $\mu \sim 7000$ cm$^2$V$^{-1}$s$^{-1}$ for the hole-doped region. In the calculation process, we used the puddle density ($n_{pd}$) as $9 \times 10^{10}$ cm$^{-2}$. On the other hand, when the same sample was placed in a vacuum chamber at a pressure of $\sim 10^{-5}$ Torr for two days, we obtained $\mu \sim 13000$ cm$^2$V$^{-1}$s$^{-1}$ for both doped regions, $n_{pd} < 3 \times 10^{10}$ cm$^{-2}$ and $V_0 \sim -0.7$ V at room $T$ (see Figs. S2b-S2d in the Supporting Information). This indicates that a graphene on hBN is seriously affected by water molecules in the air, resulting in movement of the $V_0$ with increasing $n_{pd}$ and finally leading to a reduction in mobility. Although such environmental effect could not be avoided while the Raman thermometry was performed in the ambient condition, we found that the mobility was not significantly changed from $\sim 7000$ cm$^2$V$^{-1}$s$^{-1}$ during measurements.

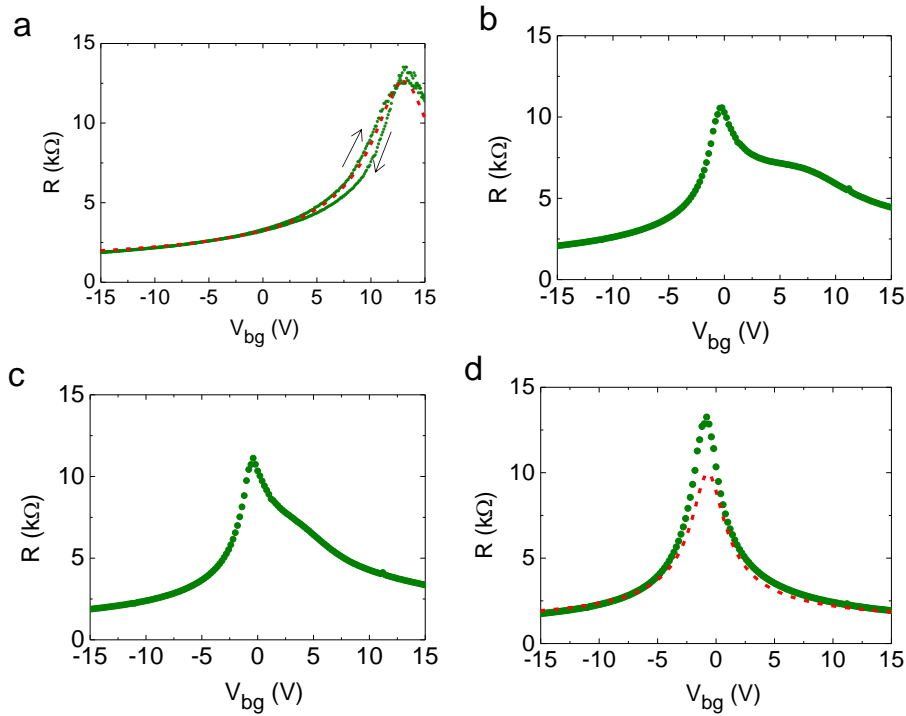

**Figure S2.** (a) $R$-$V_{bg}$ curve (scattered points) and fit-result (dashed curve) in air. Arrows indicate the gate sweep directions. $R$-$V_{bg}$ curves (b) as soon as reach to a pressure of $10^{-5}$ Torr, (c) after two hours in the vacuum and (d) after two days in the vacuum. The dashed curve in (d) is a fit-result to estimate the mobility.

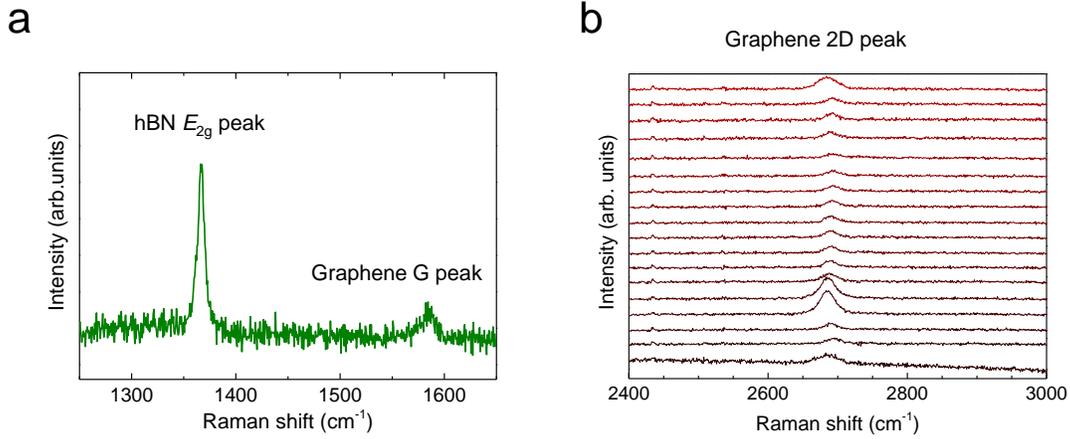

**Figure S3.** (a) Raman spectrum of the device for the $E_{2g}$ and G peaks of hBN and graphene, respectively, which was obtained from the middle of the graphene channel in Fig. 1b in the main text. (b) 2D peaks of graphene, which were obtained by scanning from source (top spectrum) to drain (bottom spectrum) direction with a 1 μm step at the middle in the width direction.

Figure S3(a) shows a Raman spectrum at the middle of the graphene channel in Figure 1b, which reveals the $E_{2g}$ phonon mode of hBN at ~1366.6 cm⁻¹ and the *G* peak of graphene at ~1583.5cm⁻¹. Figure S3(b) also shows observable 2D peaks of the graphene by scanning from the source to drain. Unfortunately, the intensities of the G and 2D peaks were not sufficiently developed to define the energy values through the graphene channel. Nevertheless, We should note that The *G* peak of graphene is sensitive to the charge doping for $n > 2 \times 10^{12}$ cm⁻² as well as *T* [2,3], which could limit its application as a thermometer because the carrier concentration along the channel is usually changed when the Fermi level changes along the channel under high-bias condition[1]. In addition, it has been known that the *T*-dependence of G peak is not linear in a *T*-range of 20-160 °C [4], which is a range of interest in this study for a 26 μm long graphene device. Meanwhile, such carrier doping effect in the hBN can be excluded due to its insulating character, thus, the energy shift of the $E_{2g}$ mode of hBN layer under high bias will provide genuine *T*-sensitive information of the device. For the 2D peak[5], although it has a carrier concentration dependence, it shows an insensitive region for $n < 10^{13}$ cm⁻². Although it was hard to get distinct graphene Raman peaks for the graphene device with $L = 26$ μm, we successfully got the G and 2D peaks for the device with $L = 5$ μm, thus, we obtained the temperature maps of graphene and hBN by the 2D and $E_{2g}$ peaks, respectively, as shown in Fig. 4 in the main text.

## Section 2. Temperature calibration, photo-induced doping effect and Raman thermometry for the $L = 26$ μm graphene device

The scattered points in Figure S4a show the energy of the $E_{2g}$ peak of the hBN as a function of $T$ using the excitation wavelength of 633 nm (red light)[6,7]. The $T$ dependence of the $E_{2g}$ phonon energy of hBN was determined by heating the graphene/hBN FET device on a hot plate. The energy of the $E_{2g}$ peak decreased with increasing $T$, which was due to an increase of the anharmonic effect of phonons with $T$ [8]. The slope of the curve was $S_{BN} = -0.0218$ cm$^{-1}$/°C. During the Raman thermometry measurements, the excitation laser power was maintained at less than 1 mW, which corresponded to a power density of approximately $0.6 \times 10^8$ mWcm$^{-2}$. Heating effects on the sample due to laser absorption were not observed. We note that the graphene/hBN heterostructure is sensitive to the photo-induced doping effect, which occurs when optical irradiation excites electrons from donor defects in the hBN.[9] We demonstrated the photo-induced doping effect with a 633 nm-wavelength laser, as shown in Figure S4b. In the beginning, the level of $R$ of the graphene FET was 6.90-7.05 kΩ at zero $V_{bg}$. Although the laser power was turned on the graphene channel, a noticeable change in the level of $R$ was not observed. In the case of a 514 nm-wavelength (green) laser, however, the photo-induced doping effect apparently changed the resistance of the graphene while $S_{BN}$ shows a similar value to that for the red laser (see Figures S4c and S4d). A previous work showed that the photo-induced doping effect for the graphene on hBN became significant when the photon energy was larger than ~2.3 eV.[9] Because the photon energies of the green and red lasers are 2.41 and 1.96 eV, respectively, our observations are consistent with the previous report on the photo-induced doping effect. The photo-induced doping effect will induce locally different doping regions along the examined graphene layer during Raman thermometry, which alters the heat dissipation role. Thus, for the Raman thermometry we adopted a red laser, which was relatively insensitive to the photo-induced doping effect on the graphene/hBN heterostructure.

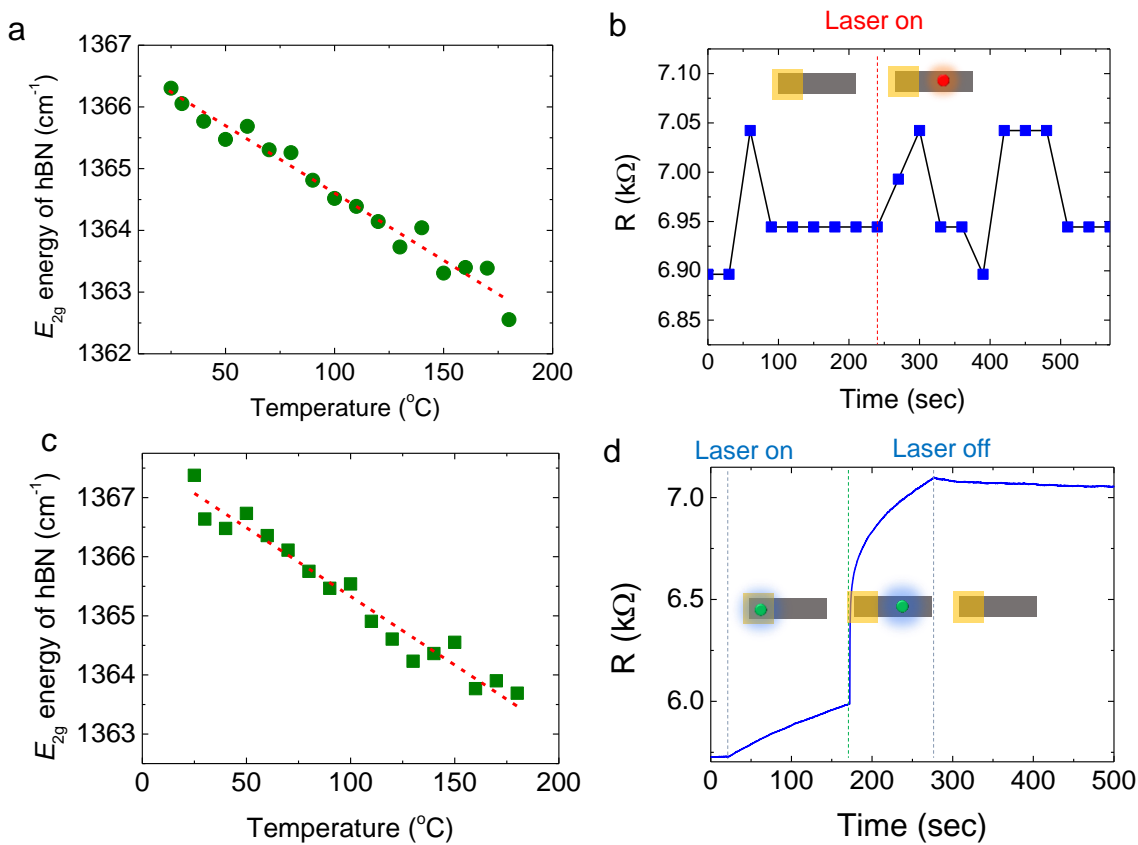

**Figure S4.** (a) Peak-energy change in the $E_{2g}$ mode of hBN as a function of temperature (scattered points) with a 633 nm-wavelength (red) laser, where the dashed line is a linear fit result with a slope of $S_{BN} = -0.0218$ cm$^{-1}$/°C. (b) Resistance change of the graphene as a function of time with and without laser irradiation on the graphene channel. (c) Peak-energy change in the $E_{2g}$ mode of hBN as a function of temperature (scattered points) with a 514 nm-wavelength (green) laser, where the dashed line is a linear fit result with a slope of $S_{BN} = -0.023$ cm$^{-1}$/°C. (d) Resistance ($R$) changes of the graphene as a function of time without and with laser irradiation on the graphene channel. The $R$ of the graphene FET was measured while the stage moved from the electrode to the graphene region with respect to the laser spot. In the beginning, $R$ of the graphene FET is constant at 5.7 k$\Omega$. When the laser power is turned on in the electrode area, the $R$ begins to increase. As soon as the laser spot enters the graphene region, the $R$ abruptly increases and the increasing rate is reduced with time. When the laser is turned off at 300 sec, the $R$ keeps a nearly constant value.

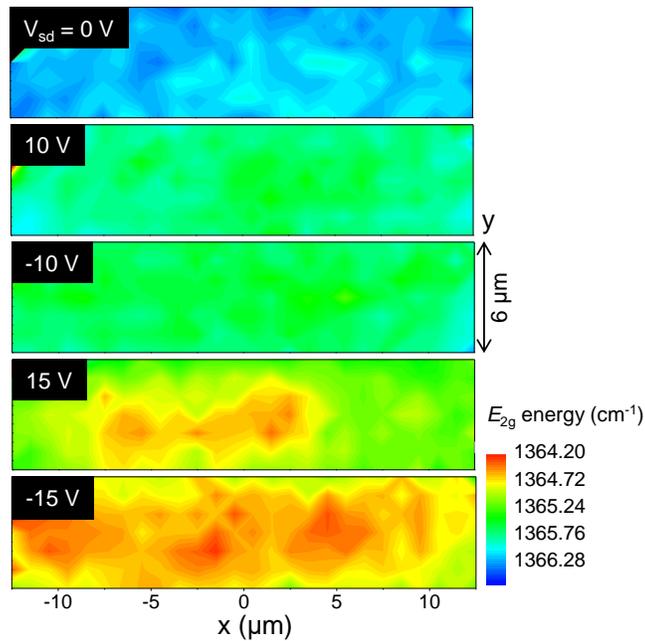

**Figure S5.** $E_{2g}$-energy maps of the hBN underneath the graphene channel for various $V_{sd}$ values. The left and right ends are drain and source, respectively. The top panel shows the $E_{2g}$-energy map of the hBN without bias voltage. Slight fluctuations in the $E_{2g}$ peak energies ($< 0.2$ cm$^{-1}$) over the mapping images could be attributed mainly to variations in the local strain in the hBN. To get a $T$-map of the hBN layer under bias, the $E_{2g}$ energy at each pixel without bias voltage at $T_0 = 22.4$ °C was used as a reference value and subtracted from the $E_{2g}$ energy at the corresponding pixel under bias. Panels from the second one to the bottom show the $E_{2g}$-energy maps of the hBN underneath the graphene channel at $V_{sd} = 10$, $-10$, 15 and $-15$ V, respectively. With increasing $V_{sd}$, the downshift of the $E_{2g}$ energy is distinguishable through the entire hBN. The $E_{2g}$-energy and temperature maps at the same $V_{sd}$ of Fig. 1d in the main panel show a slightly different distribution because the temperature at a pixel at a given $V_{sd}$ was obtained by a difference of the $E_{2g}$ energies of zero and examined biases at the corresponding pixel.

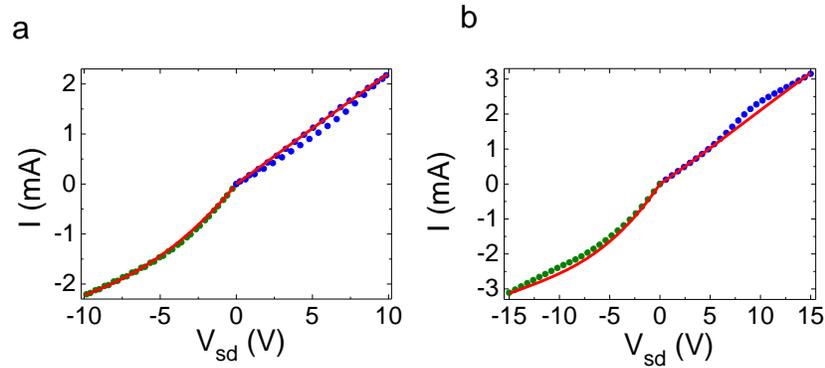

**Figure S6.** (**a**) and (**b**) Current vs. $V_{sd}$ ($I$-$V_{sd}$) curves to perform the hBN thermometry, where the Raman scans were taken under the final voltage conditions with procedures of $V_{sd} = 10$, $-10$, $15$, and $-15$ V. Scattered and solid curves indicate the experimental and calculated results, respectively.

## Section 3. Raman thermometry for $L = 5$ μm graphene device

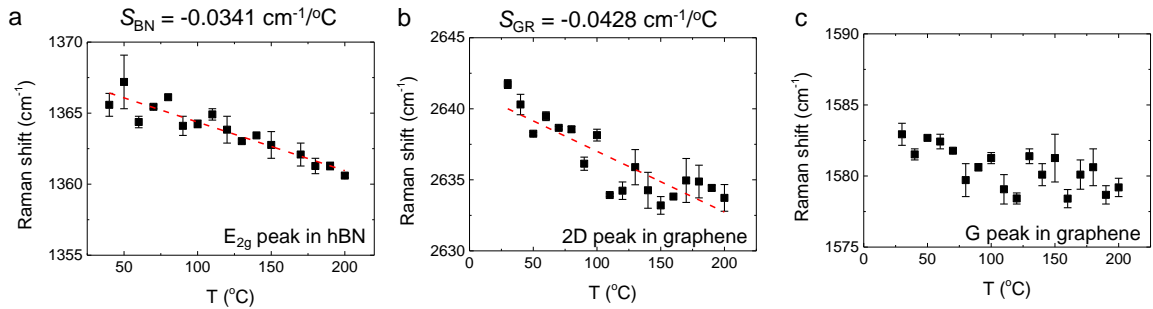

**Figure S7.** Peak-energy changes in (a) $E_{2g}$ mode of hBN, (b) 2D mode and (c) G mode of graphene as a function of temperature (scattered points), where dashed lines in (a) and (b) are linear fit results with a slope of $S_{BN} = -0.0341$ cm$^{-1}$/°C and $S_{GR} = -0.0428$ cm$^{-1}$/°C, respectively. For the G mode of graphene in (c), it is hard to find a temperature dependence, thus the G mode was not suitable for the Raman thermometry in our case.

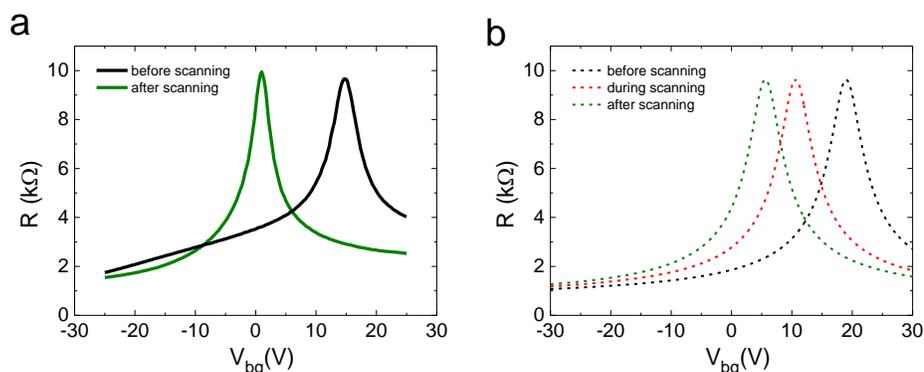

**Figure S8.** (a) $R$-$V_{bg}$ curves before and after Raman scanning at $V_{sd} = 8.5$ V, which shows a shift of the charge neutral voltage ($V_0$) after the scanning. (b) Calculation results to perform the electro-thermal calculations. In the calculation process, rather different $V_0$ were used to get a better agreement with experiments. Here, mobility was $\mu \sim 10,000$ cm$^2$V$^{-1}$s$^{-1}$.

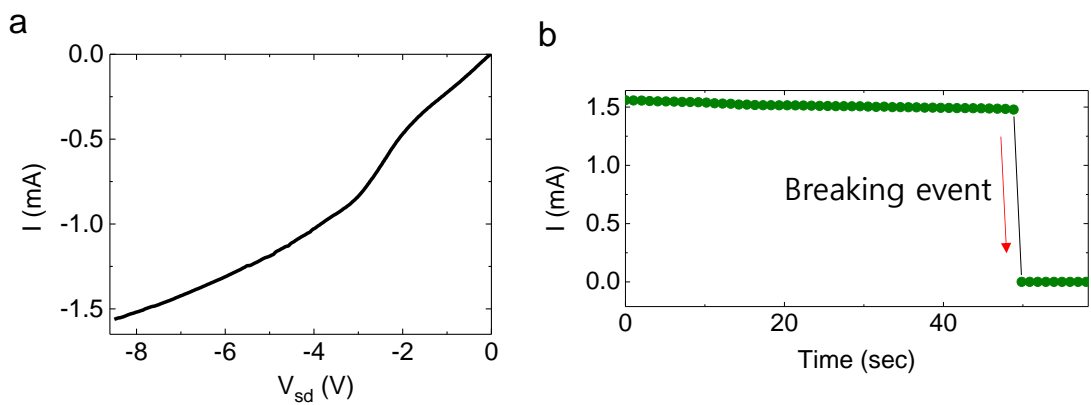

**Figure S9.** (a) $I$-$V_{sd}$ curve of $L = 5$ μm device for negative biases. (b) After reaching $V_{sd} = -8.5$ V and $I = -1.56$ mA, it shows a breaking-down event in 50 sec. The top image of Fig. 4a in the main text is an AFM image after breaking-down event. It shows a broken-graphene region near source where a hot spot is located with a negative $V_{sd}$ value in a hole-doped region.

**References for Supplementary Information**